\documentclass[a4paper,UKenglish,cleveref, autoref, thm-restate]{lipics-v2021}

\usepackage[T1]{fontenc}
\usepackage[usenames,dvipsnames]{xcolor}

\def\QQ{\mathbb{Q}}
\def\ZZ{\mathbb{Z}}
\def\CC{\mathbb{C}}
\def\NN{\mathbb{N}}

\def\RR{\mathbb{R}}

\usepackage{xspace}
\usepackage{listings}

\definecolor{keywordcolor}{rgb}{0.7, 0.1, 0.1}   
\definecolor{commentcolor}{rgb}{0.4, 0.4, 0.4}   
\definecolor{symbolcolor}{rgb}{0, 0, 0.8}    
\definecolor{sortcolor}{rgb}{0.1, 0.5, 0.1}      
		
\nolinenumbers
			
\newcommand{\lean}[1]{\lstinline{#1}\xspace} 

\title{Fermat's Last Theorem for regular primes (short paper)}

\author{Alex J. Best}{King's College London \and \url{https://alexjbest.github.io/} }{alex.j.best@gmail.com}{https://orcid.org/0000-0002-5741-674X}{}

\author{Christopher Birkbeck}{University of East Anglia \and \url{https://cdbirkbeck.wixsite.com/website}}{c.birkbeck@uea.ac.uk}{https://orcid.org/0000-0002-7546-9028}{}

\author{Riccardo Brasca}{Université Paris Cité \and \url{https://webusers.imj-prg.fr/~riccardo.brasca/}}{riccardo.brasca@gmail.com}{https://orcid.org/0000-0002-0491-7241}{}

\author{Eric Rodriguez Boidi}{King's College London}{eric.rodriguez-boidi@kcl.ac.uk}{https://orcid.org/0000-0002-0507-627X}{}

\authorrunning{A.\,J. Best, C. Birkbeck, R. Brasca, and E. Rodriguez}

\Copyright{Alexander J. Best, Christopher Birkbeck, Riccardo Brasca and Eric Rodriguez Boidi}

\ccsdesc[500]{General and reference~Verification}
\ccsdesc[300]{Computing methodologies~Representation of mathematical objects}
\ccsdesc[300]{Mathematics of computing~Mathematical software}

\keywords{Fermat's Last Theorem, Cyclotomic fields, Interactive theorem proving, Lean}

\supplement{The code described is available in a separate repository to mathlib}
\supplementdetails[linktext={https://github.com/leanprover-community/flt-regular}]{Software}{https://github.com/leanprover-community/flt-regular}

\acknowledgements{We thank the \emph{mathlib} community for a lot of useful discussions around our project. We especially thank Ruben Van de Velde for having formalised in Lean a proof of Fermat's Last Theorem in the case $n = 3$. AB. was supported by NWO Vidi grant 639.032.613}

\EventEditors{Adam Naumowicz and Ren\'{e} Thiemann}
\EventNoEds{2}
\EventLongTitle{14th International Conference on Interactive Theorem Proving (ITP 2023)}
\EventShortTitle{ITP 2023}
\EventAcronym{ITP}
\EventYear{2023}
\EventDate{July 31 to August 4, 2023}
\EventLocation{Bia\l{}ystok, Poland}
\EventLogo{}
\SeriesVolume{268}
\ArticleNo{36}

\category{Short Paper}

\begin{document}

\maketitle

\begin{abstract}
We formalise the proof of the first case of Fermat's Last Theorem for regular primes using the \emph{Lean} theorem prover and its mathematical library \emph{mathlib}. This is an important 19th century result that motivated the development of modern algebraic number theory. Besides explaining the mathematics behind this result, we analyze in this paper the difficulties we faced in the formalisation process and how we solved them. For example, we had to deal with a diamond about characteristic zero fields and problems arising from multiple nested coercions related to number fields. We also explain how we integrated our work to \emph{mathlib}.
\end{abstract}

\section{Introduction}

Fermat's Last Theorem states that for $n \geq 3$, the equation $x^n+y^n=z^n$ has no solutions with $x,y,z \in \ZZ$ and $xyz \ne 0$. This question remained unsolved for $300$ years until the eventual proof of this was completed by Andrew Wiles and Richard Taylor \cite{wiles, taylor-wiles} in 1994. This proof requires a great deal of mathematical machinery in order to study deep connections between number theory, algebra, geometry and analysis and its formalisation is currently out of reach. However, certain special cases of this theorem were already known long before Wiles' work. First of all, it's easy to prove that we can restrict to the case where the exponent $n$ is an odd prime $p$. Moreover, Kummer proved in $1847$ that

\begin{theorem}[Kummer]
Let $p$ be a regular (odd) prime. Then $x^p+y^p=z^p$ has no solutions with $x,y,z \in \ZZ$ and $xyz \ne 0$.
\end{theorem}
Here \textit{regular} means that $p$ does not divide the class number of the cyclotomic field $\QQ(\zeta_p)$, where $\zeta_p$ is a primitive $p$-th root of unity (i.e. $\zeta_p^p=1$ and $\zeta_p^k \neq 1$ for any $0 < k < p$). For example, the only irregular primes less that $100$ are $37$, $59$ and $67$. It is worth noting that currently there is no proof that there are infinitely many regular primes, but it is expected that roughly $60\%$ of primes are regular.

In this short paper we report on ongoing work, the \lean{flt\_regular} project, to formalise Kummer's proof using the Lean theorem prover \cite{lean}. We build upon the \textit{mathlib} library \cite{mathlib} is a large library of formalized mathematics for Lean. This library contains, for example, the definition of a number field and that of its ring of integers. More specifically for this work, the definitions and basic lemmas about class groups and class numbers have already been formalised \cite{class_group}, including that the class number of a number field is finite, a nontrivial fact that is required to even define regular primes properly. 

Kummer's proof can be split into two cases, depending on if $p$ is allowed to divide $xyz$ in the theorem, or not. These are known in the field as the \emph{first case} and the \emph{second case}:

\begin{theorem}[Case I]\label{thm:case_one}
	Let $p$ be a regular (odd) prime. Then $x^p+y^p=z^p$ has no solutions with $x,y,z \in \ZZ$ and $\gcd(xyz,p)=1$.
\end{theorem}

Case II then changes $\gcd(xyz,p)=1$ to $\gcd(xyz,p)=p$, and together the two cases imply Kummer's theorem. While the proofs of the first and second case are broadly similar, and use many of the same techniques, results, and ideas, the proof of the second case uses more delicate results about units in cyclotomic fields (including one known as Kummer's lemma), and is therefore more difficult to formalise, even though many underlying results are the same. The formalisation of the second case is work in progress, so we will focus on the first case from now on.

These results are by now viewed as classical results in algebraic number theory and are covered in a number of works, we have followed the standard reference \cite{washington} for the most part. We also made use of a \emph{blueprint}, an informal document included with the project covering the formalization targets in sufficient detail that the formalization progress could be tracked against it.

For large parts of this project the formalisation process and the process of adding the results to \emph{mathlib} were done almost in parallel, see Section \ref{mathlib} below for more details.

\section{Cyclotomic fields}

The formalisation begins with the definition of cyclotomic fields and of more general of cyclotomic extensions. This is necessary to even define regular primes, and also appear in the first step of the high level overview of Kummer's proof. The basic mathematical idea is to work in the field $\QQ(\zeta_p)$, that is the field obtained by adding to $\QQ$ a primitive $p$-th root of unity (in $\CC$ say). In such a field the left-hand-side of Fermat's equation can be written as
$$
x^p + y^p = (x+y)(x+\zeta_py)\cdots(x+y\zeta_p^{p-1}).
$$
Using this, one can deduce information about the rational, and integral, solutions of Fermat's equation.
The arithmetic properties of $\QQ(\zeta_p)$ (such as how integral elements decompose as products of prime elements) are much more complicated than those of $\QQ$. Studying such questions is the main subject of algebraic number theory. Fortunately for us, \emph{mathlib} already contains many of the basic definitions the we will need, such as algebras, number fields, rings of integers and class groups. Unfortunately, \emph{mathlib} did not contain examples of non-trivial number fields, so these definitions also served as a good test of the existing API.

Let $A$ and $B$ be commutative rings. For an $A$-algebra $B$ and a set $S$ of positive natural numbers, we say that $B$ is a $S$-cyclotomic extension, if for every $n \in S$ there exists a primitive $n$-th root of unity in $B$ and moreover that $B$ is generated over $A$ by the $n$-th roots of unity (for $n \in S$). We hence define a class \lean{is_cyclotomic_extension}, now part of \emph{mathlib}.

\begin{lstlisting}
@[mk_iff] class is_cyclotomic_extension : Prop :=
(exists_prim_root {n : ℕ+} (ha : n ∈ S) : ∃ r : B, is_primitive_root r n)
(adjoin_roots : ∀ (x : B), x ∈ adjoin A { b : B | ∃ n : ℕ+, n ∈ S ∧
    b ^ (n : ℕ) = 1 })
\end{lstlisting}

The choice of working with $n$ in $\NN+$ rather than in $\NN$ is motivated by the fact that, even if it requires us to insert certain coercions (for example to say that $n$ is prime), the $0$-th cyclotomic extension is not well behaved and theorems have a neater statement when that possibility is excluded.

We then define \lean{cyclotomic_field n K}, where $n$ is as above and $K$ a field, and we prove that the corresponding field extension is an instance of the \lean{is_cyclotomic_extension} class. To be precise, we define \lean{cyclotomic_field n K} as the splitting field over $K$ of the $n$-cyclotomic polynomial.

\begin{lstlisting}
@[derive [field, algebra K, inhabited]]
def cyclotomic_field : Type w := (cyclotomic n K).splitting_field    
\end{lstlisting}
Here, the \lean{derive} attribute makes the \lean{field}, \lean{algebra K} and \lean{inhabited} instances from \lean{splitting_field} apply to \lean{cyclotomic_field}. The \lean{is\_cyclotomic\_extension} instance must then be proved manually.

Mathematically, using a predicate in the way we use the class \lean{is_cyclotomic_extension} is uncommon, as one usually only works with the specific example \lean{cyclotomic_field n K} (and indeed all $n$-th cyclotomic extensions of a field $K$ are isomorphic if $n \neq 0$ in $K$), but having a characteristic predicate is essential in the formalisation process, for example to state that subextensions of a given cyclotomic extension generated by roots of unity are still cyclotomic, and to be able to apply lemmas to them. We prove several results about cyclotomic extensions and importantly, we prove that if $S$ is finite and $K$ is a number field, then any $S$-cyclotomic extension of $K$ is again a number field. This allows us to define the usual cyclotomic number fields, such as $\QQ(\zeta_n)$. 

After the setup defining such fields is done, the main task is then to prove that the ring of integers of $\QQ(\zeta_n)$ is $\ZZ[\zeta_n]$. In general, for a field $K$, its ring of integers $\mathcal{O}_K$ is the set of elements of $K$ that are roots of a monic polynomial with coefficients in $\ZZ$. In particular, the inclusion $\ZZ[\zeta_n] \subseteq \mathcal{O}_{\QQ[\zeta_n]}$ is clear, but equality is a nontrivial result that is specific to cyclotomic extensions, and we proved it only when $n = p^k$ is a power of a prime. Proving only this case is sufficient for us, indeed to prove Kummer's theorem only the case $n=p$ is needed, and in addition the general case makes use of this result for the prime-power case. This lemma is therefore a natural candidate for inclusion in a library that aims to include results in as much generality as possible, i.e. \emph{mathlib} in the case of Lean code.
To prove that $\mathcal{O}_{\QQ(\zeta_{p^k})} = \ZZ[\zeta_{p^k}]$ we had to add a considerable amount of mathematics to \emph{mathlib}, and this was the first significant milestone of the project. This required expanding the existing API for the norm and trace of elements of a number field, defining the discriminant of a number field and proving results relating discriminants to bases of rings of integers, etc., all material that would appear in a first course on algebraic number theory. Amongst the required results, we proved that, for $k > 0$ and $p > 2$ a prime, the discriminant of $\QQ(\zeta_{p^k})$ is $(-1)^{\varphi(p^k)/2} p^{p ^ {k - 1}((p-1)  k -1) },$ with $\varphi$ the Euler's totient function. Moreover, for $p =2$ and $k > 1$ the same formula holds. In \emph{mathlib} we can use the fact that as \emph{natural numbers}, we have by convention, that $1/2=0$ and $0-1=0$. This allows us to give a simple formula for the discriminant that applies in all cases:

\begin{lstlisting}
lemma discr_prime_pow {p : ℕ+} {k : ℕ} {K L : Type*} {ζ : L} [field K]
  [field L] [algebra K L] [hcycl : is_cyclotomic_extension {p ^ k} K L] 
  [hp : fact (p : ℕ).prime] (hζ : is_primitive_root ζ ↑(p ^ k)) 
  (hirr : irreducible (cyclotomic (↑(p ^ k) : ℕ) K)) :
  discr K (hζ.power_basis K).basis = (-1) ^ (((p ^ k : ℕ).totient) / 2) * 
    p ^ ((p : ℕ) ^ (k - 1) * ((p - 1) * k - 1))
\end{lstlisting}

Note that here $K$ and $L$ are only assumed to be fields (so, for example, they could have characteristic $p$), which is why we need the additional assumptions, that hold if $K = \QQ$ and $L = \QQ(\zeta_{p^k})$.

\section{About the proof of case I}
One issue that we often encountered came from a typeclass diamond resulting from multiple inheritance paths when working with a field of characteristic zero (see \cite{10.1007/978-3-030-51054-1_1} for more on how this sort of issue arises and is resolved). Our issue arises as \lean{cyclotomic\_field n K} is endowed with the instance \lean{algebra K (cyclotomic\_field n K)}, but if $K = \QQ$, then there is another instance \lean{algebra ℚ (cyclotomic\_field n ℚ)}, coming from the fact that \lean{cyclotomic\_field n ℚ} is a characteristic zero field, and hence a $\QQ$-algebra. These two $\QQ$-algebra structures were propositionally, but not definitionally, equal.
This caused some friction when using results stated via the more general instance but Lean finds the one resulting from characteristic zero. However, we were able to resolve this issue by changing the way that \lean{splitting_field} is defined.

Previously, these instances were lifted from a base field to the splitting field by direct induction, and this gave us no definitional control of the field of this structure (specifically, the \lean{qsmul} and \lean{rat_cast} fields in \lean{field (splitting_field f)}). The fix for this was to lift every field individually and put them together later, so that we can control these crucial definitional equalities. As we are lifting to a quotient, we need to take care that these operations are well defined, and this led to the introduction of \lean{distrib_smul}: a typeclass carrying the action of one type on another weak enough that the ``obvious'' map $\QQ \times K \to K$ satisfies it, but strong enough to guarantee that lifting this to the map $\QQ \times K[X] / (p(X)) \to K[X] / (p(X))$ is well defined.

Having developed the cyclotomic field framework we then moved to proving the technical number theoretic lemmas which are required in the proof of case I. These involve the careful study of units in the rings of integers of cyclotomic fields as well as certain ideals in these rings. Before describing the necessary lemmas, let us highlight a recurring issue when dealing with units.

Consider the following situation. Let $R$ be an integral domain and $K$ its field of fractions. Given a unit $r \in R^\times$, we may want to think of $r$ as an element of $R^\times$, $R$ or $K$. In \emph{mathlib} these are all different types so we need coercions maps between them. Now for $r$ as an element of $R^\times$ and $K$, we can easily work with its inverse, i.e, we can consider $r^{-1}$, but this is not possible when considered as an element $R$, since $R$ is only a ring, so in general elements don't have a multiplicative inverse, but when coerced one step further to elements of the field $K$ we are once again able to define a well defined inverse function. These issues arise often when working with ideals, which are submodules of $R$ but when the proofs require one to use units in several places. Note that it is not clear how to set up \lean{simp} lemmas that normalise elements, since sometimes we want to move from $R$ to $K$ and sometimes from $R$ to $R^\times$. The solution to this is to have simple lemmas relating the images of $r^{-1}$ in $R$ and $K$.

\begin{lstlisting}
lemma coe_coe_inv (u : Rˣ) : ((u : R) : K)⁻¹ = ((u⁻¹ : Rˣ) : R)
\end{lstlisting}

As an example of where this is used we have the following lemma, where $K$ will denote $\QQ(\zeta_p)$ and $R=\ZZ[\zeta_p]$. We will also denote $\zeta_p$ by $\zeta$; note that \texttt{h$\zeta$.unit'} is the same as $\zeta$, but considered in the units $R^\times$.

\begin{lemma}
Let $p \ne 2$ be a prime. Then every unit $u \in \ZZ[\zeta_p]^\times$ can be written as $u=x \zeta_p^n$ for some $n \in \ZZ$ and $x \in \ZZ[\zeta_p]^\times$ such that $x \in \RR$.
\end{lemma}
\begin{lstlisting}
lemma unit_lemma_gal_conj (h : p ≠ 2) (hp : (p : ℕ).prime) (u : Rˣ)
  (hζ : is_primitive_root ζ p) :
  ∃ (x : Rˣ) (n : ℤ), is_gal_conj_real p (x : K) ∧ 
    (u : R) = x * (hζ.unit' ^ n : Rˣ)
\end{lstlisting}
Here the integer $n$ cannot be supposed to be in $\NN$, so $x$ must be an element of a group (namely $R^{\times}$), to allow integer-valued powers. On the other hand, the existence of $x \in K$ is not enough, so we both need $R$ and $R^{\times}$. Finally, the Galois group acts on $K$, so we really need the three different types to state the lemma cleanly. Note also that we state the informal condition that $x \in \RR$ to \lean{is_gal_conj_real p x}, which says that $x$ is fixed under complex conjugation, where complex conjugation is thought of as an element of $\mathrm{Gal}(\QQ(\zeta_p)/\QQ)$. In particular, we can avoid the non-canonical coercion into the real numbers used implicitly in standard proofs by reformulating what it means to be ``real'' in this setting. 

This situation is perhaps not yet fully satisfactory, as manually rewriting to convert between the same element coerced into different types forces us to work at a lower level than we would when discussing the material informally. Another solution such as more automation may be better in the long term.

For brevity, we will not list all of the lemmas required to prove case I, but full details can be found on our project blueprint here: \url{https://leanprover-community.github.io/flt-regular/}. The final result we prove is

\begin{lstlisting}
theorem caseI {a b c : ℤ} {p : ℕ} [fact p.prime] (h : is_regular_prime p) 
  (caseI : ¬ ↑p ∣ a * b * c) :
  a ^ p + b ^ p ≠ c ^ p
\end{lstlisting}
and a full sorry-free proof can be found in \url{https://github.com/leanprover-community/flt-regular/}. We note that the case of $p=3$ can be done without the tools we have formalised here. In fact, in this case, the result was formalised by Ruben van de Velde at \url{https://github.com/Ruben-VandeVelde/flt} using elementary methods.

We end this section with the definition of \lean{is_regular_prime}. Here \lean{is_regular\_number} says that a positive integer $n$ is regular if $n$ is coprime to the size of the class group of $\QQ(\zeta_n)$ from which we define \lean{is_regular_prime} as the condition that a prime number is regular. We currently have a proof that $p=2$ is a regular prime, but in general proving that a certain prime is regular (with our current definition) requires us to compute the relevant class number, which in general is difficult to do in \emph{mathlib}, this is something that will be addressed in the proof of case II, which will relate being regular to a more easily checkable condition (factorizations of certain Bernoulli numbers).
Explicit calculations of class numbers of quadratic fields have been formalized \cite{10.1145/3573105.3575682}, and while the cyclotomic fields we use here are in general not quadratic, some of the same techniques may be of use when calculating class numbers of cyclotomic fields directly.

\begin{lstlisting}
def is_regular_number [hn : fact (0 < n)] : Prop :=
n.coprime
  (card (class_group (ring_of_integers (cyclotomic_field ⟨n, hn.out⟩ ℚ))))
\end{lstlisting}
 
\section{Integration to \emph{mathlib}} \label{mathlib}
While working on a mathematical formalization such as this one, newly introduced material is often not stable. For instance, new definitions are often changed as working with them reveals deficiencies, and proof strategies are factored out into common lemmas or abstractions when they are recognized after being seen several times.
This means that when working on such a project new material is initially not ready for use outside of the project as it may change radically to suit the needs of the project.
Nevertheless contributing material to a large library is a way to ensure continued maintenance of the code, especially when the upstream library changes. Thus contribution to a library may be desirable when the code is sufficiently mature, despite the fact that adding such material to a large library requires external review and may take time.

One slightly unusual aspect of our work is that we tried to include our results in \emph{mathlib} almost in real time, keeping the two projects closely in sync. This is in contrast to many other similar projects where first the main theorem is formalised in its entirety and then one begins the process of adding the results to \emph{mathlib}, which usually results in a great deal of modifications to the original code. For example the \emph{Perfectoid Project} \cite{perfectoid} and the \emph{Liquid Tensor Experiment} \cite{LTE} both have huge \lean{for\_mathlib} folders with a lot of formalised mathematics that in principle is supposed to be integrated into \emph{mathlib}, but the code does not yet have the required standards of quality and generality. This state is often reached as authors do not have the time required to polish the code to the standard required and open PRs to contribute it. Developing against a library with little to no backwards compatibility maintained such as \emph{mathlib} then means that maintaining the code of a large project can be a painful job. For example, the \emph{Perfectoid Project} is essentially stuck to a very old version of \emph{mathlib}, and updating it to the latest version is a nearly impossible task.

Our approach is to have a folder \lean{ready\_for\_mathlib} where we put as much as results as possible, opening PRs immediately. Even if this means sometimes proving certain results in unneeded generality, we think this is the best strategy for a medium-sized project as this one. Moreover, this also implies that our code is up to the standard of \emph{mathlib}, that are usually very high. This kept the size of the \lean{flt\_regular} project relatively small, but for example the whole folder \lean{number\_theory/cyclotomic} (that is around 2000 lines of code) in \emph{mathlib} was written as a byproduct of our work. One other side effect is that updating \emph{mathlib} is a rather easy process: we are indeed using the latest version and we plan to keep doing so. In practice we opened (and had accepted) more than 110 PRs, in various areas of mathematics, ranging from linear algebra to number theory. A partial history of the PRs opened can be seen at at \hyperlink{https://github.com/leanprover-community/flt-regular/wiki}{https://github.com/leanprover-community/flt-regular/wiki}.

In order for large ecosystems of formal proofs, such as Lean's mathlib and surrounding libraries, to continue to scale and cover a significant portion of graduate level mathematics it seems more automation will be necessary to ease the contribution and organisational burden. The fields of program repair and automated refactoring (and more specifically the burgeoning field of proof repair \cite{ringer}, with more emphasis on mathematical proofs) provide a model for what should be possible and useful.
For instance when a large library that a project such as  \lean{flt\_regular} depends on is updated, an automated summarisation of changes and required modifications (or even automated patch creation and application) would reduce the manual effort keeping up to date with a rapidly changing upstream.
Automated style fixers and code improvers (to simplify the process of golfing and generalizing working proofs or to avoid anti-patterns) would reduce the effort required to contribute functional but less mature proofs to a standard library.
Automation to help more specifically with moving code between libraries, by situating it correctly and updating local import paths would also improve the workflow.

\section{Future work}

The next step in our work will be to give a full proof of Fermat's Last Theorem in the regular case. The main obstacle here is to prove Kummer's lemma:

\begin{theorem}\label{thm:Kummers_lemma}
	Let $p$ be a regular prime and let $u \in \ZZ[\zeta_p]^\times$. If $u \equiv a \mod p$ for some $a \in \ZZ$, then there exists $v \in \ZZ[\zeta_p]^\times$ such that $u=v^p$.
\end{theorem}

There are several ways to prove this lemma, with modern approaches using class field theory. For our purposes this approach would take is too far afield from our final goal. The first step will be to use an alternative definition of regular prime, which instead of asking that $p$ does not divide the class number of $\QQ(\zeta_p)$ asks that $p$ does not divide the numerator of certain Bernoulli numbers. This definition also has the added benefit that it is easy to check that a prime is regular, since Bernoulli numbers are easy to compute (and this is already in \emph{mathlib}). This then leaves the task of checking that these definitions are equivalent, which can be done without using class field theory, but will still require significant work. Following classical proofs the main obstacle in proving this equivalence of definitions (and Kummer's lemma) will be the need to understand the image in the $p$-adic completion of $K$ the logarithm of certain units. Amongst other things the final proof of case II will require the formalisation of $p$-adic completions of number fields and their extensions. Furthermore we will require analytic results for $p$-adic logarithms and their links to Bernoulli numbers.

Several results relating Bernoulli numbers to values of $p$-adic $L$-functions have been formalized by Narayanan \cite{narayanan2023formalization}, these results may form a nontrivial part of the formalization of case II, depending on the approach taken.

\bibliography{biblio}

\end{document}